\documentclass[aps,prx,reprint,superscriptaddress,showpacs]{revtex4-1}
\usepackage{graphicx}
\usepackage{amssymb}
\usepackage{color}
\usepackage{amsmath}
\usepackage{ulem}
\usepackage{mathrsfs}

\newcommand{\be}{\begin{equation}}
\newcommand{\ee}{\end{equation}}

\begin{document}

\title{Quantum-critical scaling properties of the two-dimensional random-singlet state}

\author{Lu Liu}
\affiliation{Beijing National Laboratory for Condensed Matter Physics, Institute of Physics, Chinese Academy of Sciences, Beijing 100190, China}

\author{Wenan Guo}
\email{waguo@bnu.edu.cn}
\affiliation{Department of Physics, Beijing Normal University, Beijing 100875, China}
\affiliation{Beijing Computational Science Research Center, Beijing 100193, China}

\author{Anders W. Sandvik}
\email{sandvik@bu.edu}
\affiliation{Department of Physics, Boston University, 590 Commonwealth Avenue, Boston, Massachusetts 02215, USA}
\affiliation{Beijing National Laboratory for Condensed Matter Physics, Institute of Physics, Chinese Academy of Sciences, Beijing 100190, China}

\begin{abstract}
  We use QMC simulations to study effects of disorder on the $S=1/2$
   Heisenberg model with exchange constant $J$ on the square lattice
   supplemented by multispin interactions $Q$. It was found recently
   [L. Lu et al., Phys. Rev. X 8, 041040 (2018)] that the ground state
   of this $J$-$Q$ model with random couplings undergoes a quantum
   phase transition from the N\'eel state into a randomness-induced
   spin-liquid-like state that is a close analogue to the well known
   random-singlet (RS) state of the random Heisenberg chain. The 2D RS
   state arises from spinons localized at topological defects. The
   interacting spinons form a critical state with mean spin-spin
   correlations decaying with distance $r$ as $r^{-2}$, as in the
   1D RS state. The dynamic exponent $z \ge 2$, varying continuously
   with the model parameters. We here further investigate the properties
   of the RS state in the $J$-$Q$ model with random $Q$ couplings.
   We study the temperature dependence of the specific heat and various
   susceptibilities for large enough systems to reach the thermodynamic
   limit and also analyze the size dependence of the critical magnetic
   order parameter and its susceptibility in the ground state. For all
   these quantities, we find consistency with conventional
   quantum-critical scaling when the condition implied by the $r^{-2}$
   form of the spin correlations is imposed. All quantities can be
   explained by the same value of the dynamic exponent $z$ at fixed
   model parameters. We argue that the RS state identified in the $J$-$Q$
   model corresponds to a generic renormalization group fixed point
   that can be reached in many quantum magnets with random couplings,
   and may already have been observed experimentally.
\end{abstract}

\date{\today}

\maketitle

\section{Introduction}
\label{sec:intro}

Effects of disorder and impurities play an important role in quantum many-body physics, not only because of their perturbing effects on uniform
systems but also because the interplay of disorder and quantum fluctuations can lead to completely different states. Prominent examples are Anderson
localization of electrons \cite{anderson} and its proposed generalization to many-body localization in interacting systems \cite{fleishman,altshuler,basko,gornyi,pal}.
Another interesting and well known state induced by disorder is the random singlet (RS) state in one dimensional (1D) spin chains. The RS state is the fixed
point of the simple but powerful strong-disorder renormalization group (SDRG) method \cite{skma1,skma2,dsfisher,rnbhatt}, where pairs of strongly coupled spins
are successively decimated. The fixed point is an infinite-randomness fixed point (IRFP), where the effective strength of the randomness (the width of the
coupling distribution on a logarithmic scale) increases to infinity without bounds under the renormalization group (RG) flow.

The 1D RS state resulting from the SDRG method for a system such as the $S=1/2$ antiferromagnetic Heisenberg chain with random couplings is a special 'frozen'
configuration of random singlet spin pairs (valence bonds), and this state accurately represents the true ground state, which is a superposition of valence-bond
coverings. The distance between two spins in a singlet is typically short, but there are rare instances of very long valance bonds that cause the mean spin
correlations to decay with distance as $r^{-2}$, while the typical correlations decay exponentially. A striking property of the IRFPs is that the dynamic exponent
$z$ is infinite, leading to unconventional dynamical scaling properties. In this paper we will present a detailed numerical characterization of a two-dimensional
(2D) state of $S=1/2$ spins that is in many respects similar to the 1D RS state, though it has finite, but large, dynamic exponent.

\subsection{Ground states of 2D random quantum magnets}
\label{sub:2dstates}

\begin{figure}[t]
\centering
\includegraphics[width=70mm, clip]{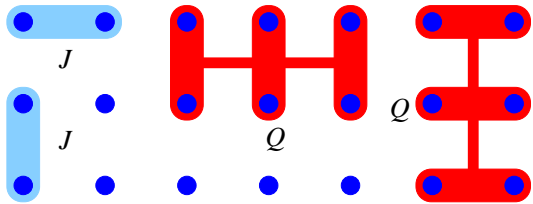}\\
\caption{Illustration of the terms in the 2D $J$-$Q$ model studied in this paper. Sites on the square lattice are shown as blue dots. The light blue bars
represent the $J$ terms while the groups of three connected red bars represent the products of three singlet projectors constituting each $Q$ terms.}
\label{terms}
\end{figure}

The 1D RS phase is broadly realized in random spin chains with different symmetries, not only with the fully rotationally invariant Heisenberg (XXX)
interactions but also in anisotropic XXZ chains and in the transverse-field Ising model (TFIM) \cite{dsfisher}. In gapless host systems such as the $S=1/2$
XXX and XX chains, even infinitesimal randomness drives the system asymptotically into the RS phase, while in uniform gapped systems (e.g., the integer-$S$
Heisenberg chains with Haldane gaps) a critical disorder (width of the coupling distribution) is required \cite{hyman97}. In spontaneously dimerized
systems with random couplings, such as the frustrated $J_1$-$J_2$ Heisenberg chain \cite{lavarelo13} or the $J$-$Q$ chain \cite{shu16} (a 1D variant of
the model studied in this paper, which is illustrated in Fig.~\ref{terms}), domain walls with localized spinons form for any disorder strength and those
spinons can form the IRFP-RS state in some cases \cite{shu16} (and other states formed by the spinons have also been proposed \cite{lavarelo13}).

In two-dimensional (2D) systems, although IRFPs have been identified in TFIMs \cite{kovacs10}, no convincing evidence of such a state has been reported
in 2D quantum magnets with SU(2) invariant interactions \cite{melin00,damle00,motrunich00,lin03,lin06}; only a spurious IRFP was
pointed out in Ref.~\onlinecite{laumann12}. Unlike the 1D case, an RS state is not obtained in the 2D Heisenberg model with bond randomness if
all couplings remain antiferromagnetic (i.e., in the absence of frustration), with the N\'eel ground state surviving any strength of such disorder
\cite{laflorencie06}. The robustness of the 2D N\'eel state is also exemplified by the site diluted system, which remains ordered all the way to the percolation
point \cite{sandvik02a}. Interesting quantum states can be observed at the percolation points in site and bond diluted
systems \cite{sandvik02b,vajk02,yu05,sandvik06,wang10}, and in some systems with two spins per unit cell in the clean limit, gapless disordered Griffiths
phases have been observed \cite{ma15}. SDRG studies of various Heisenberg systems have found what appear to be finite-randomness fixed points
\cite{melin00,damle00,motrunich00,lin03,lin06}. We note that the SDRG method can be applied generally, but if the decimation process does not flow to an
IRFP, it is not clear whether the resulting state has any bearing on the true ground state. In particular, it is not known whether a finite-disorder
fixed point \cite{iyer12} is correctly reproduced by the method.

Some frustrated quantum spin systems with random couplings have been shown to host spin-glass states, where, in analogy with classical spin models, there is no
spatial order but the individual spin expectation values $\langle {\bf S}_i \rangle$ are frozen in time \cite{bray80}. However, while some 2D $S=1/2$ TFIMs
with short-range interactions definitely exhibit spin glass behavior \cite{usadel86,rieger94}, no convincing case of SU(2) short-range interactions producing
spin glasses have been presented. Analytical calculations with long-range interactions or some extension of the symmetry group, e.g., SU($N$), have demonstrated
spin glasses \cite{bray80,sachdev93,read95}, but the relevance of such solutions to short-range interacting SU($2$) spins is not clear. Nevertheless,
it has been proposed that models such as the 2D Heisenberg with random nearest-neighbor couplings (with mixed signs of the couplings to introduce frustration)
can host spin glass phases, e.g., based on numerical exact diagonalization (ED) of small systems \cite{oitmaa01}. Extrapolating the properties based on ED
to the thermodynamic limit is challenging, however, and other interpretations are also possible.

Alternatives to spin glass states have been proposed in the generic context of random frustrated interactions. In long-range interacting models a ``spin fluid''
phase with $\langle {\bf S}_i \rangle=0$ and unusual dynamic properties similar to the marginal Fermi liquid was proposed \cite{sachdev93}, but again these
properties do not necessarily carry over to short-range interacting $S=1/2$ spins. Numerical studies of $S=1/2$ systems with short-range interactions on various
lattices have suggested states such as the valence-bond glass \cite{tarzia08,singh10} and the randomness-induced quantum spin liquid
\cite{watanabe14,kawamura14,shimokawa15,uematsu17,uematsu18,kawamura19,uematsu19,hanqing19,huanda20}. The latter series of works includes also the square-lattice
$J_1$-$J_2$ Heisenberg model \cite{uematsu18}, for which a spin glass ground state had previously been proposed \cite{oitmaa01}. Terms such as ``valence-bond
glass'' and ``randomness-induced spin liquid'' do not yet have completely well defined meanings in terms of unique system properties related to specific RG
fixed points in two space dimensions. It nevertheless appears that 2D states exist that have no spatial order, are not spin glasses (i.e.,
$\langle {\bf S}_i \rangle=0$), but are gapless and in some ways similar to the 1D RS state. We will here characterize such a state in detail,
based on a model amenable to quantum Monte Carlo (QMC) simulations (Fig.~\ref{terms}).

\subsection{Routes to the 2D random-singlet state}

Recently, two independent works using different techniques \cite{kimchiprx,luliu} pointed out that a state similar to the 1D RS state can be obtained by starting
from a 2D valence-bond solid (VBS) state, i.e., a gapped state in which lattice symmetries are spontaneously broken due to the formation of a singlet pattern. Due
to the Imry-Ma argument for disorder in systems with discrete symmetry breaking \cite{imry75}, extended to quantum systems, the VBS pattern is destroyed for any
finite amount of coupling disorder, and in such a ``broken VBS'' topological defects along with unpaired spins (before interactions are considered) form localized
spinons. These spinons, which are related to the spinons in clean VBS systems associated with deconfined quantum-critical points \cite{levin04,Shao}, are coupled
to each other via effective interactions mediated by the host system, and the question is then what kind of magnetic subsystem they give rise to.

Kimchi et al.~\cite{kimchiprx} argued that a state similar to the 1D RS state can
form if the host system is frustrated, e.g., in the Heisenberg model on the triangular lattice, though eventually, at the lowest energy scales, a spin
glass state may form (though no actual evidence was presented for a spin glass). They also suggested that an unfrustrated (bipartite) host system does not support
the RS state but leads to antiferromagnetically ordered spinons. In our previous work with collaborators \cite{luliu}, we considered a ``designer'' $J$-$Q$ model,
illustrated in Fig.~\ref{terms}, and demonstrated that its VBS state in the presence of disorder leads to domains and localized spinons, but there were no signs of
the spinons forming long-range order. Instead, it was found that the spin correlations always decay with distance as $r^{-2}$, as in the 1D RS state, and
quantum-critical scaling properties at finite temperature $T$ were also identified. All scaling properties were explained with a dynamic exponent which takes the
value $z=2$ at the transition from the N\'eel state and grows inside the RS phase. Values up to $z \approx 7$ were found in the $J$-$Q$ model, and there is likely
no upper bound in principle in an extended model space. We also presented arguments based on spinon pairing and domain-wall mediated interactions as to why no ordering takes place. The identified RS scaling
properties then correspond to an asymptotic RG fixed point. We also suggested that this RS fixed point found in a system without traditional frustration
(but with other forms of competition between different interactions)
may be the same as the one identified by Kimchi et al.~in frustrated models \cite{kimchiprx}, and which was also argued to be realized in some materials
\cite{kimchinatcom} that had previously been regarded as quantum spin liquids \cite{li15} or quantum spin glasses \cite{ma18}.

The existence of a generic disorder-induced spin liquid had previously been argued based on numerical ED studies by Kawamura and collaborators
\cite{watanabe14,kawamura14,shimokawa15,uematsu17,uematsu18}. The method only allows access to small system sizes, and it is difficult to interpret the
results and extract long-distance and low-energy properties. ED results had previously been interpreted in terms of a spin glass \cite{oitmaa01}.
The state studied in these ED works may actually be equivalent, in the RG sense, to the RS state discussed in Refs.~\onlinecite{kimchiprx,luliu}.
It was also recently proposed that the starting point of the randomness-perturbed VBS state, where there is still substantial short-range VBS
correlations and the spinons form a dilute subsystem, can be regarded as a weak-disorder variant of the state originating from strong randomness in
frustrated systems \cite{kawamura19}. There may potentially be no phase transition separating the weak and strong disorder cases, and instead one can think
of continuously shrinking the VBS domains discussed in Refs.~\cite{kimchiprx,luliu} until the domains become so small that it would be meaningless to consider them
as domains. Instead the random valence-bond configurations may become more akin to a valence-bond glass \cite{tarzia08,singh10}. The crucial point here is
whether the localized spinons of the weak-disorder regime survive also at strong disorder. Indeed, also in the ED studies, ``orphan spins'' were identified
that may be those spinons \cite{watanabe14,kawamura14,shimokawa15,uematsu17,uematsu18,kawamura19}, though their role in forming the RS state was less clear
because of the small system sizes and smaller distance between the orphan spins than between the spinons in the limit of large VBS domains \cite{luliu}.
Anderson localization of singlets had also been proposed as the mechanism responsible for the formation of RS state \cite{shimokawa15}.

A very recent study based on a semi-classical approach gives support to the notion that spin glasses in classical frustrated systems transform into states
akin to the 1D RS state when strong quantum effects are included \cite{dey20}, and that the mechanism of starting from the VBS states to understand such
a 2D RS state \cite{kimchiprx,luliu} also is valid for systems that are not VBS ordered in the clean limit, e.g., the frustrated Heisenberg model on the
triangular lattice (which has three-sublattice antiferromagnetic order for all spin values $S$).

If the RS state identified in the $J$-$Q$ model indeed
corresponds to the same fixed point as in the frustrated systems, for which further evidence based on density matrix renormalization group (DMRG) calculations
has been presented recently \cite{huanda20}, then this fixed point, as well as the phase transition into the N\'eel state, can be fully characterized in
detail using QMC simulations. This is the {\it designer Hamiltonian} approach \cite{kaul13} of studying specifically tailored lattice models that do
not necessarily correspond to microscopic descriptions of specific materials but enables reliable calculations of universal properties.

\subsection{Paper motivation and outline}

The motivation for the present work is to test quantum-critical scaling laws in the random $J$-$Q$ model to a higher degree than what was previously
done in Ref.~\onlinecite{luliu}. On the basis of such results we can make definite predictions for the RS phase that can be tested also in other models with
different methods, as well as in materials. Because of the very significant computational efforts required, we focus on a single point inside the RS phase
and study very large systems down to ultra-low temperatures, in order to reliably obtain the asymptotic temperature dependence of the specific
heat (which was not calculated in our previous work) and various susceptibilities (for which we report much improved results). For the same set of model
parameters, we also study the ground state of systems on smaller lattice sizes (lengths) $L$ and carry out finite-size scaling analysis. The $T>0$,
$L \to \infty$ scaling properties are fully consistent with the finite-$L$, $T \to 0$ finite-size behaviors with the same dynamic exponent $z \approx 4.0$,
thus supporting a finite-disorder fixed point with formally conventional scaling behavior but with a large dynamic exponent $z$ and spin correlations
decaying universally as $r^{-2}$ (for which we also show additional evidence here).

The paper is organized as follows: In Sec.~\ref{sec:randomjq}, we introduce the random-$Q$ $J$-$Q$ model. In Sec.~\ref{sec:specficheat}, we present scaling
results for the specific heat, which we obtain by analyzing the internal energy density. In Sec.~\ref{sec:susc}, we show results for the temperature dependent
local, uniform and staggered susceptibilities, extracting the asymptotic low-temperature forms in the thermodynamic limit. In Sec.~\ref{sec:t0} we discuss
the staggered structure factor and susceptibility for smaller systems in the ground state, from which we extract an independent estimate of the dynamic
exponent $z$. We conclude in Sec.~\ref{sec:diss} with a discussion of the relevance of our findings to the present status of the 2D RS state, including
possible experimental realizations.

\section{2D random $J$-$Q$ model and simulation method}
\label{sec:randomjq}

The $J$-$Q$ model on a 2D square lattice has nearest-neighbor antiferromagnetic exchange $J$ along with multi-spin interactions of strength
$Q$ \cite{sandvik07} expressed using products of singlet projectors. We here study the variant in which the $Q$ terms are products of three singlet
projectors, also called the $J$-$Q_3$ model \cite{lou09}, with the Hamiltonian is defined as
\begin{equation}
\label{jqham}
H = -J \sum \limits_{\langle ij\rangle} P_{ij} -
Q \hskip-3mm \sum \limits_{\langle ijklmn\rangle} \hskip-2.5mm P_{ij}P_{kl}P_{mn},
\end{equation}
where $P_{ij}$ is the singlet projector for two $S=1/2$ spins at sites $i$ and $j$,
\begin{equation}
P_{ij} = \frac{1}{4} - {\bf S}_i \cdot {\bf S}_j.
\end{equation}
In the Hamiltonian Eq.~(\ref{jqham}),
${\langle ij\rangle}$ indicates nearest-neighbor sites, and the index pairs $ij$, $kl$, and $mn$ in ${\langle ijklmn\rangle}$ form three parallel
bonds on horizontal or vertical columns, as illustrated in Fig.~\ref{terms}. The summations are over all nearest neighbors for the $J$ terms and
all vertical and horizontal columns for the $Q$ terms, so that the Hamiltonian does not break any lattice symmetries.

In the clean system, the $J$ and $Q$ terms compete with each other, individually having different ground states, leading to a phase transition
from the standard N\'eel ordered ground state for small $Q/J$ to the spontaneously dimerized columnar VBS state for large $Q/J$ \cite{sandvik07,lou09,Shao}.
Such a N\'eel--VBS transition appears to realize the deconfined quantum criticality (DQC) scenario \cite{Senthil1,Senthil2}, a continuous order--order transition
beyond the conventional Landau-Ginzburg-Wilson paradigm according to which the transition should be generically first-order. While we will not discuss the quantum
phase transition in the present work, the N\'eel--VBS transition may be considered the clean limit of the transition between the N\'eel state and the
RS state forming due to disorder in either the $J$ or $Q$ terms \cite{luliu}.

To study effects of disorder, we here introduce randomness in the $Q$ terms, choosing the strength of each of them to be either $0$ or $2Q$ at
random with equal probability for the two cases; then the mean strength $\bar Q$ of these interactions is still $Q$, which we take as our tuning parameter.
We define the energy unit with $J=1$. For this random-$Q$ model, in our previous work we detected a quantum phase transition from the N\'eel phase
to the RS phase upon increasing $Q$ \cite{luliu}, with critical point located at $Q_c \approx 1.2$. At the critical point, a dynamic exponent consistent
with $z=2$ was found. The entire RS phase is a critical state without magnetic long range order, with the mean spin-spin correlations decaying with distance
$r$ as $r^{-2}$. We confirmed this behavior for several points inside the RS phase, not only for the model with random bimodal $Q$ but also with bimodal
$J$ and with continuous distributions of the couplings. We found consistently that $z \approx 2$ on the AFM--RS phase boundary (most
likely $z=2$ exactly), and $z$ grows monotonically on moving deeper into the RS phase. Thus, while $z$ can become large, this RS state does not correspond
to an IFRP as the 1D variant \cite{skma1,skma2,dsfisher}. In Sec.~\ref{sec:diss} we will further discuss why the term RS state is still an appropriate
way to characterize the state.

Our main aim here is to study several different quantities at high statistical precision, in order to test quantum-critical scaling laws in which the exponent $z$
appears. In some quantities $z$ is the only exponent controlling the asymptotic behavior, while in other quantities the exponent governing the decay of the spin
correlation function also enters. By testing for consistent scaling to high precision among several different quantities, we will solidify the claim that conventional
quantum-critical scaling applies, with a single well-defined parameter-dependent dynamic exponent and with a universal $r^{-2}$ form of the spin correlations.
These calculations are very demanding of computational resources, and we therefore only focus on a single point, $Q=2$, in the phase diagram of the bimodal random-$Q$
$J$-$Q$ model. This point is well away from the phase boundary to the N\'eel phase and should represent the generic behavior with only $z$ varying as we move
throughout the RS phase.

We use the stochastic series expansion (SSE) QMC method with efficient loop updates \cite{sse1,sse2}, which produces exact results within statistical errors. We
average all results over more than $10^4$ disorder realizations for the final production runs. In the calculations of finite-$T$ properties, reported below in
Secs.~\ref{sec:specficheat} and \ref{sec:susc}, we start at a relatively high temperature, where the simulation equilibrates easily, and gradually lower $T$,
keeping the disorder realization fixed and continuing the simulation from the last configuration generated at the previous temperature. We accumulate averages
of the observables of interest at each temperature on the chosen $T$ grid (in some cases with a varying distance between the $T$ points, as will be described
further below). The number of SSE configuration updates performed at each step is large enough to ensure that the system stays in equilibrium and that
sufficient statistics can be collected for each disorder realization at each $T$. Typically we use $10^3$ steps for equilibration and $10^3$ for accumulating
averages at each $T$. As we will see, the statistical fluctuations of the final disorder-averaged quantities are dominated by the intrinsic sample-to-sample
variations, even with this rather modest number of steps per temperature.

To obtain results representing the thermodynamic limit at low temperatures, we need to use very large system sizes, up to $L=256$, and we anneal down to temperatures
as low as $T_0/J=1/150$ for the largest system size. In the calculations aimed at ground state properties, the temperature has to be much lower in order to
avoid finite-temperature effects, and we have only gone up to size $L=48$ in these calculations.

\section{Specific heat}
\label{sec:specficheat}

The specific heat $C$ plays an important role in experimental studies of quantum magnetism, providing thermodynamic information about the system.
In QMC simulations, the specific heat at low temperatures is in general difficult to compute precisely, as the relative error bars grow when $C$
decreases. We did not consider $C$ in our previous work on the random $J$-$Q$ model, though it has been computed recently for other variants of the
$J$-$Q$ model and provided experimentally useful bench-mark results \cite{guo20}.

To analyze the asymptotic low-$T$ behavior, we have found it better to not compute $C$ directly in the simulations, but to instead
analyze the internal energy $E(T)$, which is produced by the SSE method with very high precision even at low $T$. By fitting a suitable function
to $E(T)$, we can deduce the low-temperature properties of the specific heat in the RS phase.

At a quantum-critical point, the specific heat in general scales with temperature as \cite{fisher89}
\be
C(T) \propto T^{D/z},
\label{cpscale}
\ee
where $D$ is the spatial dimensionality of the system; here $D=2$. Our hypothesis is that this form, as well as other standard quantum-critical scaling
laws \cite{fisher89}, holds also inside the RS phase. As already mentioned, our previous work on other quantities showed that $z \ge 2$ varies.

\begin{figure}[t]
\includegraphics[width=75mm, clip]{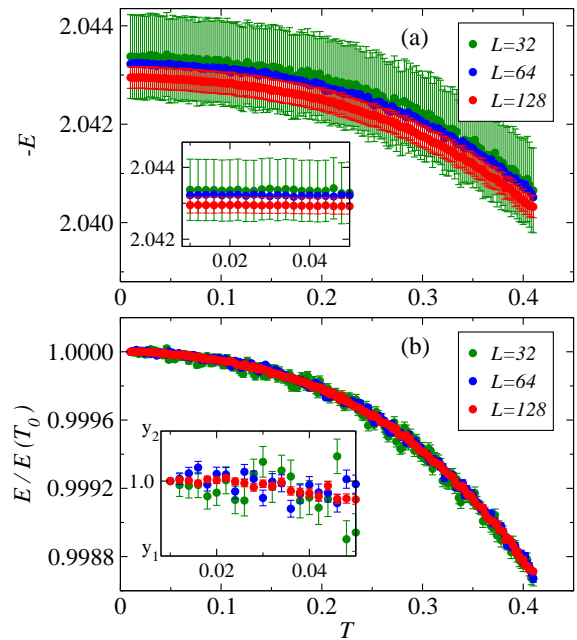}\\
\caption{Internal energy density vs temperature for system sizes $L=32$, $64$, and $128$, averaged over $10^3$ disorder realizations. The raw data are
  shown in (a) and the results normalized by $E(T_0)$, with $T_0=0.01$, are shown in (b). The insets show magnified graphs of the low-$T$ data.
  In (a) we do not show the error bars for $L=64$ in order to avoid clutter (the size of these errors is between those for $L=32$ and $L=124$).
  In the inset of $b$, we have marked the lowest and highest points on the vertical axis by $y_1=0.99994$ and $y_2=1.00004$.}
\label{cpall}
\end{figure}

From the definition of the specific heat, $C(T)=\partial E(T)/\partial T$, we obtain the expected scaling form of the internal energy $E(T)$ per spin
according to Eq.~(\ref{cpscale}),
\be
E(T) = E_0+ A T^{D/z+1},
\label{escale}
\ee
where $E_0$ is the ground state energy density.
We show SSE results for $E(T)$ in Fig.~\ref{cpall}(a) for three different system sizes, $L=32,64$ and $128$. The annealing process used in the simulations
started at $T=0.4$, where the system equilibrates easily (recall that we set $J=1$ and report all quantities in units of energy with this convention, and the
mean of the bimodal $Q$ value is $\bar Q=2$). Within the error bars we do not observe any finite-size effects, though certainly we do expect some. It can be
seen clearly that the errors are highly correlated, with the points for each $L$ forming a curve that is much smoother than what would be expected
with independent statistical errors with standard deviation given by the error bars. Clearly, this substantial covariance is due to the fact that the same
randomness realizations are used at all temperatures, and the overall fluctuations are primarily due to sample-sample fluctuations.

\begin{figure}[t]
\includegraphics[width=75mm, clip]{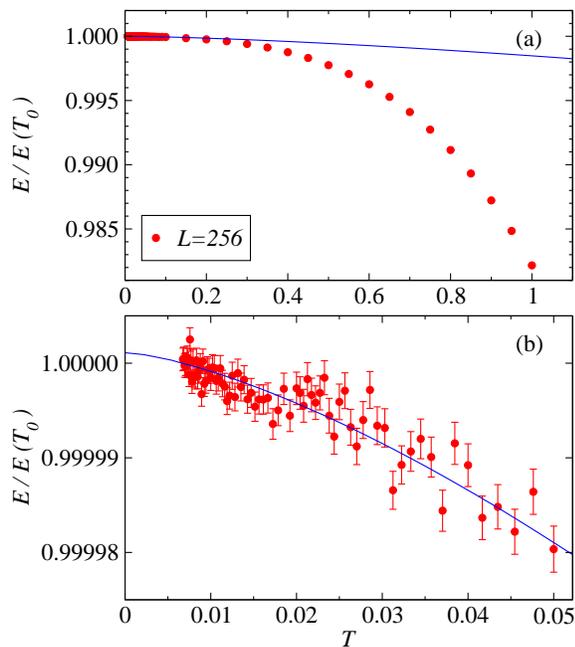}
\caption{Normalized energy $E(T)/E(T_0)$ obtained by averaging over more than $10^4$ disorder realizations with $L=256$ and $T_0=1/150$.
The blue curves show a fit of the data up to $T_{\rm max}=0.1$ to Eq.~(\ref{escale2}). The exponent $\alpha$ is graphed versus $T_{\rm max}$ in Fig.~\ref{expon}.
Panel (a) shows the whole range of date and (b) focuses on the results for $T$ up to $0.1$.}
\label{cp}
\end{figure}

Since we are only interested in the exponent governing the asymptotic low-temperature form, we write the expected low-$T$ form Eq.~(\ref{escale}) as
\be
E(T) = E(T_0)+ A (T^{\alpha}-T_0^{\alpha}),
\label{escale2}
\ee
where $T_0$ is the lowest temperature studied for a given system size and we have defined the exponent $\alpha=2/z+1$. We can then study the behavior
of the ratio $E(T)/E(T_0)$, for which we expect the asymptotic form
\be
\frac{E(T)}{E(T_0)} = 1 + A_0 (T^{\alpha}-T_0^{\alpha}),
\label{escale3}
\ee
where now we only have two free parameters; the important exponent $\alpha$ and the unimportant factor $A_0$. Having eliminated the constant $E_0$
present before the normalization in Eq.~(\ref{escale2}) helps to stabilize the results for $\alpha$ in data fits. Moreover, as shown in Fig.~\ref{cpall}(b),
the error bars of the normalized ratio (computed using bootstrapping) are much smaller, by orders of magnitude, than those of the raw data,
and the remaining fluctuations have very little covariance. By definition, the size dependence now should also be much smaller, since all values
$E(T_0)=1$ if we use the same $T_0$ for all system sizes. We do not see any size dependence within the error bars in the temperature window considered
in Fig.~\ref{cpall}(b), and the remaining error bars after the normalization are smaller for the largest system, due to self-averaging. The number of
disorder samples ($10^3$) and the number of SSE updates for each temperature ($10^3$ for equilibration and the same number for evaluating expectation
values) were identical for all the system sizes in this figure.

In order to confidently study even lower temperatures, for our final analysis we simulated more than $10^4$ disorder realizations for system size $L=256$
down to $T_0=1/150$. In order to be able to reach such low temperatures on a fine $T$ grid, while keeping the total time of annealing from the highest to
lowest $T$ reasonable, we take the following strategy: After equilibrating a simulation for a given disorder realization at the highest $T$ (here
$T=1$), we decrease $T$ by a constant step $\Delta T$ and continue the simulation from the last SSE configuration of the previous temperature. After reaching
a certain low $T$ (here $T=0.1$) where it becomes necessary to decrease the step size, in order to have sufficient resolution to investigate the low-$T$
behavior and also to make sure that equilibrium is properly maintained, we modify the annealing procedure and start to increase the inverse temperature
$\beta$ by a constant step $\Delta\beta$. We thus obtain enough points in the low-$T$ region and the results are reliable.

Based on the above tests for smaller systems with $T_0=1/100$, we are confident that the larger system size with the modified annealing procedure gives us
results valid as the thermodynamic limit. Normalized results are presented in Fig.~\ref{cp}, with all results up to
$T=1$ shown in Fig.~\ref{cp}(a) and with Fig.~\ref{cp}(b) focusing on the lower temperatures, where, as we will see below, the asymptotic behavior holds
without detectable corrections. 
The error bars are further
reduced because of the larger system size and larger number of disorder samples; $10^4$ versus $10^3$.

\begin{figure}[t]
\includegraphics[width=75mm, clip]{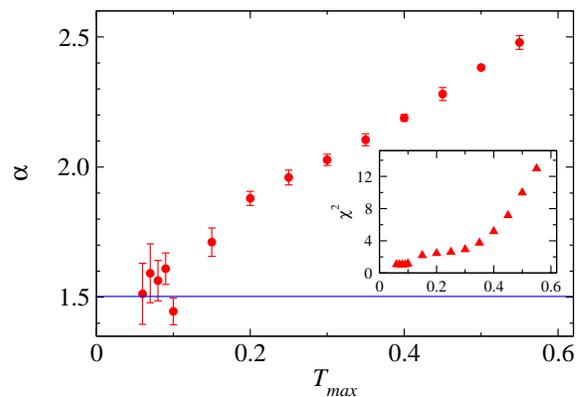}
\caption{Exponent $\alpha=2/z+1$ in the power law fit, Eq.~(\ref{escale3}), of the normalized energy as a function of the highest temperature $T_{\rm max}$
included. The data are from Fig.~\ref{cp}. Based on the value of the reduced $\chi^2$ values, which are shown in the inset, the fits are statistically acceptable
for $T_{\rm max} \le 0.1$. The result $\alpha=1.503(6)$ from the susceptibility fits in Sec.~\ref{sec:susc} is shown as the blue line.}
\label{expon}
\end{figure}

When fitting to Eq.~(\ref{escale3}), the first assumption of course is that $T_0$ is sufficiently low for the asymptotic behavior to have set in with very small
corrections. Even if $T_0$ is sufficiently low, the corrections will be important above some higher temperature. We therefore carry out fits with data up
to a maximum temperature $T_{\rm max}$ and monitor the goodness of the fit defined as $\chi^2$ per degree of freedom. Fig.~\ref{expon} shows the resulting
exponent $\alpha$ versus $T_{\rm max}$ and also the $\chi^2$ values of the fits. We can clearly observe how the fit gradually improves as $T_{\rm max}$ is
reduced, finally becoming statistically acceptable at $T_{\rm max} \approx 0.1$. The exponent $\alpha=2/z+1$ converges to a value close to $1.5$, i.e., the
specific heat exponent $2/z \approx 0.5$ in Eq.~(\ref{cpscale}) and $z \approx 4.0$. In the next section we will discuss the scaling behaviors of the uniform
susceptibility $\chi_{\rm u}$ and the local susceptibility $\chi_{\rm loc}$, which both should scale with the exponent $2/z-1$. The value of $\alpha=2/z+1$ obtained
from these quantities is consistent with the value from the specific heat and has smaller error bars, $\alpha = 1.503(6)$. We show this value as the blue
line in Fig.~\ref{expon}.

\section{Susceptibilities at finite temperature}
\label{sec:susc}

In this section, we focus on the temperature dependence of three different susceptibilities in the thermodynamic limit. As in the previous section,
we have studied several different system sizes and in the final analysis used the largest size, $L=256$, for which our convergence tests show that
the results are free from finite-size effects down to the lowest temperatures in the simulations. The data to be presented below are from the same
simulations as the energy results in Fig.~\ref{cp}.

The uniform magnetic susceptibility is defined as
\begin{equation}
\chi_{\rm u} = \frac{1}{TN}\langle m_z^2\rangle,
\label{chiudef}
\end{equation}
with the total magnetization
\begin{equation}
m_z = \sum_{i=1}^N S^z_i.
\end{equation}
The local susceptibility at location ${\bf x}$ is defined by the Kubo integral
\begin{equation}
\chi_{\rm loc}({\bf x}) = \int_0^{1/T} d\tau \langle S^z_{\bf x}(\tau)S^z_{\bf x}(0)\rangle,
\label{chilocdef}
\end{equation}
where $S^z_{\bf x}(\tau)$ is the standard imaginary-time dependent spin accessible in QMC simulations. In Ref.~\cite{luliu} we showed examples of the
local variations in $\chi_{\rm loc}({\bf x})$ and how regions of large response correlate to the presence of spinons and domain walls.
Here we will just study the average over ${\bf x}$ and refer to this quantity as $\chi_{\rm loc}$. We also calculate the staggered susceptibility
defined by the Kubo integral
\be
\chi_{\rm s}=\frac{1}{N}\int_0^{1/T} d\tau \langle m_s(\tau)m_s(0)\rangle,
\label{chistagdef}
\ee
where $m_{\rm s}$ is the staggered magnetization,
\begin{equation}
m_{\rm s} = \sum_{i=1}^N (-1)^{x_i+y_i}{S_i^z}.
\end{equation}

In a quantum critical system in the thermodynamic limit, the low-temperature scaling properties of these observables depend on the dynamic exponent $z$,
and $\chi_{\rm loc}$ and $\chi_{\rm s}$ also depend on the exponent $\eta$ governing the asymptotic correlation function. Following
Refs.~\cite{fisher89} and \cite{chubukov94}, the uniform susceptibility should take the form
\begin{equation}
\chi_{\rm u} \propto  T^{D/z-1},
\label{chiutdep}
\end{equation}
with $D=2$. We note that this form is identical to the form of $C/T$ according to Eq.~(\ref{cpscale}).

The local susceptibility, Eq.~(\ref{chilocdef}), is the integral of the spin correlation in the imaginary time direction. Translating
the results for disordered boson systems by Fisher et al.~\cite{fisher89} to spins, the mean spin-spin correlations in imaginary time at zero
spatial separation should follow the scaling law
\begin{equation}
\langle S^z_{\bf x}(\tau)S^z_{\bf x}(0)\rangle \propto \tau^{-(D+z-2+\eta)/z}.
\label{cortimedep}
\end{equation}
Making use of the previous result \cite{luliu} that the equal-time correlation function
$C(r) =\langle S^z(0)S^z(r)\rangle$ scales with distance as $r^{-2}$, the general critical form $|C(r)| \propto r^{-(D+z-2+\eta)}$
implies the relationship $\eta=2-z$ for the RS state. The time dependence in Eq.~(\ref{cortimedep}) then takes the simpler form
\begin{equation}
\langle S^z_{\bf x}(\tau)S^z_{\bf x}(0)\rangle \propto  \tau^{-2/z}.
\label{cortimedep2}
\end{equation}
 The local susceptibility (\ref{chilocdef}) should then take the following forms depending on the value of $z$;
\begin{equation}
\chi_{\rm loc} =
\left \{ \begin{array}{ll} a + b\ln{(1/T)}, & {\rm for}~ z=2, \\ cT^{2/z-1}, & {\rm for}~z>2, \end{array} \right.
\label{chiloctdep}
\end{equation}
with non-universal constants $a$, $b$, and $c$. For $z>2$, this is exactly the same as the common asymptotic form for $\chi_{\rm u}$ and $C/T$. Note again
that this form relied on the constraint imposed by the spin correlations decaying as $r^{-2}$, which can thus be implicitly tested by the local susceptibility.
In ref.~\cite{luliu} we demonstrated the logarithmic $z=2$ form of Eq.~(\ref{chiloctdep}) at the phase transition between the N\'eel and RS states. For the point
inside the RS phase studied here, we should only expect the $z > 2$ form.

The staggered susceptibility $\chi_s$ is the space-time integral of the staggered correlation function. The often used finite-size scaling
version, which we will study in Sec.~\ref{sec:t0}, is $\chi_s \propto L^{2-\eta}$, which with the RS relationship $\eta=2-z$ becomes $\chi_s \propto L^z$.
At $T>0$ in the thermodynamic limit, the integration  cut-off is given by the size $\beta=1/T$ of the system in the time dimension. Since
distances in space ($r$) and imaginary time ($\tau$) are related to each other in scaling theory by $\tau \propto r^z$, $L^z$ in the finite-size
scaling form should be replaced by $\beta$ when the space dimensions have been effectively taken to infinity.
We thus expect
\be
\chi_s \propto T^{-1}
\label{stagtdep}
\ee
for the asymptotic form of the staggered susceptibility calculated on a sufficiently large lattice. Thus, because of the specifics of the RS
state, this temperature dependent quantity does not allow for an independent test of $z$, though it implicitly provides a nice way of checking
the relationship $\eta=2-z$.

\begin{figure}[t]
\includegraphics[width=80mm, clip]{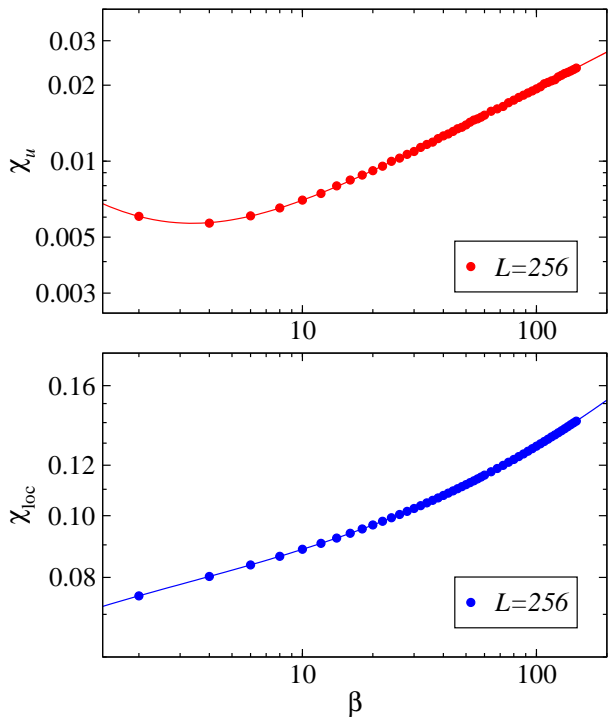}
\caption{Temperature dependence of the disorder-averaged uniform (a)  and local (b) susceptibility for systems of size $L=256$.
The curves are fits to the form $f(T)=a+bT^{-c}+dT^e$, where $c=1-2/z>0$ controls the leading (divergent) behavior. The values are
$c=0.494(9)$ for $\chi_{\rm u}$ and $c=0.500(8)$ for $\chi_{\rm loc}$. In the correction term the exponent $e$ is positive, with $e=0.97(6)$
for $\chi_{\rm u}$ and $e=0.60(2)$ for $\chi_{\rm loc}$. The error bars on the QMC data are smaller than the graph symbols.}
\label{suscep}
\end{figure}

We first discuss results for the uniform and local susceptibilities, which both should diverge asymptotically as $T^{-(1-2/z)}$, with $z>2$. As already noted
in Ref.~\onlinecite{luliu}, the corrections to the asymptotic form are much larger in $\chi_{\rm loc}$ than in $\chi_{\rm u}$. In both quantities, good
fits require the use of appropriate scaling corrections, and it was found that a constant added to the divergent form works well with both quantities.
We now have data at lower temperatures, and the statistical error are also much improved. The results for the largest system size, $L=256$, are
shown in Fig.~\ref{suscep}. Using the form $f(T)=a+bT^{-c}$ to independently fit the two data sets, we can include only the data points for $\beta=1/T \ge 20$
in the case of $\chi_{\rm u}$ and $\beta \ge 40$ in the case of $\chi_{\rm loc}$. These fits give $c=1-2/z=0.51(2)$  and $0.48(2)$ for $\chi_{\rm u}$
and $\chi_{\rm loc}$, respectively, which are consistent with each other and also with the exponent $\alpha=2/z \approx 1.5$ obtained from the
specific heat fits in Fig.~\ref{expon}

To further test the stability of the exponents obtained above, we next include a second correction term in the fitting function, $f(T)=a+bT^{-c}+dT^e$, with which
we can obtain statistically sound fits to the two susceptibilities in much larger range of temperatures; up to the highest temperature $T=1$ simulated (as
shown with the curves in Fig.~\ref{suscep}). Even though we now have two more free parameters, the larger $T$ range has a net positive effect on the statistical
precision of the leading power law in the fit. The exponents obtained are $c=1-2/z=0.494(9)$ for $\chi_{\rm u}$ and $0.500(8)$ for $\chi_{\rm loc}$, which gives us high
confidence in the exponents really being the same, and, therefore, that the form $r^{-2}$ of the spin correlation used to constrain the scaling form for $\chi_{\rm loc}$
is correct. The correction term $\propto T^{e}$ has a positive exponent $e$ (given in the caption of Fig.~\ref{suscep}) for both quantities.

There is not much variance between the two susceptibilities, and we take the average value of the two exponents as the final exponent estimate;
$1-2/z=0.497(6)$. Then the exponent governing the internal energy, discussed in the previous section, is $\alpha=2/z+1=1.503(6)$, and this value is indicated
with the blue line in Fig.~\ref{expon}. We also note here the data for smaller systems ($L=64$) and higher temperatures in the previous work \cite{luliu} gave
$1-2/z=0.60(8)$ from the scaling of the uniform susceptibility,  which also agrees with our new estimate, though the error bar of the old result is much
larger. Our best estimate of $\alpha$ corresponds to the value $z=3.98(5)$ of the dynamic exponent for the present model parameters.

\begin{figure}
\includegraphics[width=80mm, clip]{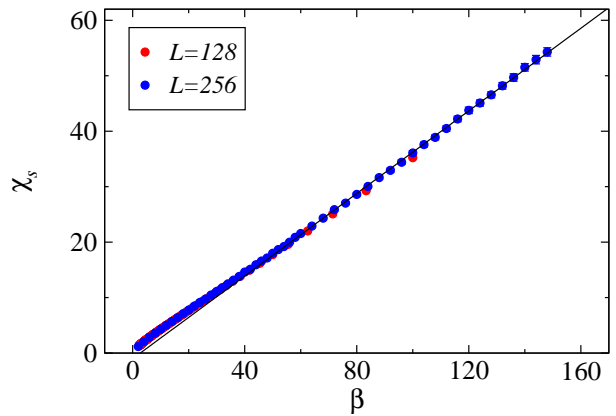}
\caption{Dependence of the staggered susceptibility on the inverse temperature for system sizes $L=128$ and $256$. The predicted linear
form, Eq.~(\ref{stagtdep}), is fully consistent with the data starting from $\beta \approx 50$, as shown with the fitted line.}
\label{scaltau}
\end{figure}

In Fig.~\ref{scaltau} we test the predicted linear form (\ref{stagtdep}) of the staggered susceptibility. We show results for two different system sizes, $L=128$ and
$L=256$, to illustrate the absence of significant finite size effects up to $\beta=100$, where we have data for both systems. The further results for $L=256$ up to
$\beta=150$ continue with the same linear trend and there is no reason to suspect significant size effects here. The results from $\beta=50$ to $150$ are fully
consistent with the linear form.

\section{Ground state finite-size scaling}
\label{sec:t0}

It is also useful to study systems at sufficiently low temperatures for computed quantities to have converged to their ground state values. The dynamic
exponent $z \approx 4.0$ estimated from the scaling behaviors analyzed in the previous sections also controls the finite-size excitation gaps in the system;
we expect the smallest gap to scale as $\Delta \propto L^{-z}$. Asymptotically, for large systems we therefore have to reach down to temperatures of order $L^{-z}$
to converge to the ground state. However, as we will show below, the gaps also have corrections that corresponds to an effective, size-dependent dynamic exponent
$z_{\rm eff}$ that is smaller than $z$ for small systems (and we will extrapolate for the asymptotic value $z$). Moreover, in principle we can use temperatures
of the form $T(L) = aL^{-b}$ with any exponent $b>0$ and still reach the ground state up to some system size if the proportionality factor $a$ is small enough. In
practice, we have here used $T = aL^{-2}$ with different prefactors to test the convergence, and found that all results are well converged to the ground state with
$a=1$ up to $L=48$. Our aim here is to use the size dependence of appropriate ground state quantities to obtain an independent estimate of the dynamic exponent,
and also to further test the $r^{-2}$ form of the spin correlation function.

An often used method to test the scaling of the gap without accessing any excited states directly is to compute the static structure factor
$S({\bf q})$ and the corresponding static susceptibility $\chi({\bf q})$ at the wave-vector ${\bf q}$ of interest. The ratio
$\Delta^*(q) = 2S({\bf q})/\chi({\bf q})$ is an upper bound to the true gap, $\Delta({\bf q}) \le \Delta^*({\bf q})$.
The bound is derived from sum rules for the dynamic structure factor and is exact for a single mode.
In general, the bound may not be very good quantitatively, but for a critical mode $\Delta^*$ is expected to have the same asymptotic scaling properties
as the gap. We refer to see Ref.~\onlinecite{wang10} for more details on this approach and an application to another disordered spin system.
We will investigate the scaling properties of the ratio $S({\bf Q})/\chi({\bf Q})$, where ${\bf Q}$ is the antiferromagnetic wave-vector.

The static structure factor is just the Fourier transform of the spin correlation function,
\be
S({\bf q})=\sum_{{\bf r}} {\rm e}^{-i{\bf q}\cdot{\bf r}}\langle S^z(0) S^z({\bf r})\rangle.
\label{sf}
\ee
We set ${\bf q}=(\pi,\pi)$ and name this structure factor $S(\pi)$. If the spin correlations decay as $r^{-2}$ in the RS state, we expect the
asymptotic size dependence to be
\be
S(\pi)=a+b\ln(L).
\label{spisize}
\ee
The corresponding staggered susceptibility $\chi_{\rm s}=\chi({\bf Q})$ was already defined in Eq.~(\ref{chistagdef}) and we mentioned its expected finite-size
scaling form when the relationship $\eta=2-z$ is imposed;
\be
\chi_s\propto L^z.
\ee
As mentioned above, for a critical mode $S({\bf Q})/\chi({\bf Q})$ is normally expected to scale as $L^{-z}$, but in the case at hand here there is a
logarithmic correction, due to Eq.~(\ref{spisize}) because of the $r^{-2}$ decay in two dimensions. We will nevertheless investigate the
ratio $\chi_s/S(\pi)$ and attempt to extract $z$ from its leading asymptotic $L^z$ form. In addition, we can also extract $z$ from just the
divergence of $\chi_{\rm s}$, which is not affected by the logarithmic correction.

\begin{figure}[t]
\includegraphics[width=80mm, clip]{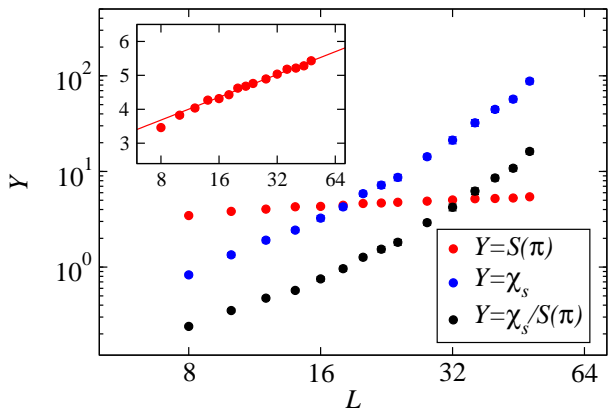}
\caption{Log-log plot showing the size dependence of $S(\pi)$, $\chi_s$, and $\chi_s/S(\pi)$ for system sizes up to $L=48$. The inset shows a log-linear plot
of $S(\pi)$, making clear the logarithmic divergence of the form Eq.~(\ref{spisize}), shown by the fitted line.}
\label{finitesize}
\end{figure}

In Fig.~\ref{finitesize} we graph the size dependence of $S(\pi)$, $\chi_s$, and $\chi_s/S(\pi)$ on a log-log scale. In the case of $S(\pi)$, the behavior looks
almost constant in the main graph, but when zooming in (the inset of Fig.~\ref{finitesize}) and graphing on a log-linear scale, the expected logarithmic divergence
is apparent. Both $\chi_s$ and $\chi_s/S(\pi)$ diverge strongly, but we do not observe a clear power-law behavior. The upward curvature on the log-log plot
implies that the effective dynamic exponent $z_{\rm eff}$ increases with increasing system size, and corrections to scaling must be considered for the available
system sizes in order to extract the asymptotic value $z_{\rm eff}(L \to \infty) \to z$.

\begin{figure}
\includegraphics[width=80mm, clip]{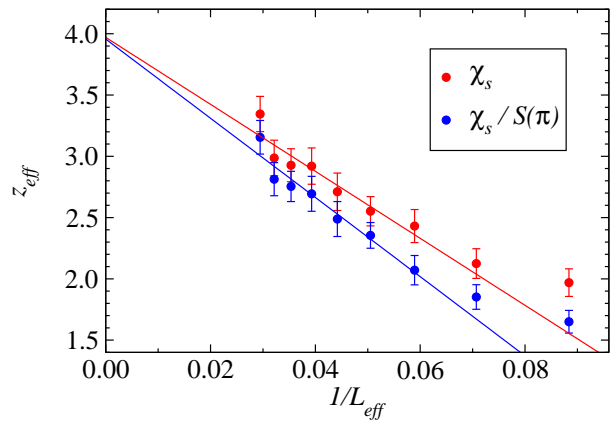}
\caption{Effective dynamic exponent defined using two system sizes, $L$ and $2L$, according to Eq.~(\ref{zeffdef}) for the quantities
$\chi_{\rm s}$ (red symbols) and $\chi_{\rm s}/S(\pi)$ (blue symbols). The effective system size is defined as $L_{\rm eff} = (2L^2)^{1/2}$.
The lines show fits giving the estimates of the extrapolated exponents $z = z_{\rm eff}(L \to \infty)$ mentioned in the text.}
\label{slope}
\end{figure}

We can define an effective dynamic exponent $z_{\rm eff}$ for a pair of systems with sizes $L$ and $2L$,
 \be
 z_{\rm eff}(L_{\rm eff})=\frac{1}{\ln(2)}[\ln(Y_{2L})-\ln(Y_L)],
\label{zeffdef}
 \ee
 where the effective system size can be defined as $L_{\rm eff}=(2L^2)^{1/2}$ (though we could also just use $L$ as the size).
 and $Y$ stands for one of the two observables; $\chi_s$ or $\chi_s/S(\pi)$.  As shown in Fig.~\ref{slope}, both of these two effective exponents converge
 to the same value, $z\approx 4$, approximately linearly with $1/L_{\rm eff}$ as system size grows. Linear fits of the form  $z_{\rm eff} = z + c /L_{\rm eff}$
 with the data starting from  $L_{\rm eff} > 16$ give $z=3.96(17)$ from $\chi_s$ and $z=3.94(14)$ from $\chi_s/s(\pi)$,  both in full agreement with our
 best determination of the dynamic exponent, $z=3.98(5)$, in the previous section. We have also fitted $\chi_s(L)$ to two power laws, which gives $z = 3.7(3)$.
 Clearly the smaller error bars when fitting to $z_{\rm eff}$ is due to the assumed linear behavior when fitting the data in Fig.~\ref{slope}.

\section{Conclusion and discussion}
\label{sec:diss}

We have studied the properties of the RS state in the 2D random-$Q$ $J$-$Q$ model, using large-scale QMC simulations to test low-temperature critical
scaling forms in the thermodynamic limit as well as finite-size scaling forms in the ground state. A salient feature of the RS state is that the spin-spin
correlations at $T=0$ decay with distance as $r^{-2}$. Formally, the exponent $2$ implies the relationship $\eta=2-z$ in scaling forms. Quantities
that do not involve $\eta$, such as the uniform and local susceptibilities, exhibit low-temperature behaviors expected in critical systems \cite{fisher89}
with a common value of the dynamic exponent $z$. Quantities with scaling forms involving $\eta$ are fully consistent with $\eta=2-z$ with the same
$z$ as that extracted from the other quantities. The high-precision calculations reported here for a single point inside the RS phase, in combination with
the less precise results for several points in Ref.~\onlinecite{luliu}, establish beyond reasonable doubt that the RS phase in the $J$-$Q$ model corresponds
to a finite-disorder critical RG fixed point, with universal $r^{-2}$ decay of the spin correlations and varying dynamic exponent $z \ge 2$.

Below, in Sec.~\ref{sub:term} we provide some further comments on the term ``2D RS state'' as it compares to the well known 1D counterpart.
In Sec.~\ref{sub:general} we discuss the possible generality of the RS state characterized here and how our findings fit into related scenarios
for randomness-induced states proposed within other approaches. In Sec.~\ref{sub:experiments} we discuss potential experimental realizations of RS
states and promising avenues to further investigate candidate materials.

\subsection{On the RS state terminology}
\label{sub:term}

The 1D RS state is an IRFP, and one may then ask whether the term 2D RS state is even appropriate for the finite-disorder fixed point we have established
here. Presumably, the RS term was coined to describe the end result of the SDRG method, which is a single valence-bond configuration with a
characteristic distribution of the length of the bonds \cite{skma1,skma2}. This single configuration is of course only a caricature of the actual
superposition ground state of a system such as the $S=1/2$ Heisenberg chain with random couplings, but it still captures the asymptotic properties
correctly in one dimension \cite{dsfisher} (perhaps up to logarithmic corrections that have been discussed recently \cite{shu16}). In other words,
the SDRG method produces a typical valence bond configuration drawn from the true ground state.

The typicality aspect of the SDRG method and the RS state is closely connected to the
IRFP nature of the fixed point. Conversely, one might not expect the SDRG method applied to 2D systems with finite-disorder fixed points to produce
completely representative results of the true ground state, which is just another way of saying that the SDRG method is reliable only when
it flows to an IRFP. It has still been argued that useful approximations can be obtained. For instance, in the early work by Bhatt and Lee \cite{rnbhatt},
the logarithmic width of the coupling distribution was found to not grow under the SDRG procedure in the case of 2D dilute randomly located moments, and the
susceptibility was found to diverge as $T^{-\alpha}$ with $\alpha <1$. These are not signatures of an IRFP, and the latter behavior can be interpreted as
a finite value of the dynamic exponent $z$ according to Eq.~(\ref{chiutdep}). Nevertheless, the SDRG method produced a set of frozen singlets, e.g.,
an RS state, and the results are believed to capture the essential physics of the system studied.

We have here not discussed the microscopic mechanisms underlying the RS state, which were described in detail in Ref.~\onlinecite{luliu}, and also within
a different theoretical framework in Ref.~\onlinecite{kimchiprx}. The essence is that the mechanism of interacting spinons in a background of VBS domains is a
close analogy to the 1D case. The analogy is the closest when the 1D RS state is constructed starting from a spontaneously dimerized system (e.g., the $J$-$Q$
chain for large $Q/J$), instead of a critical system like the standard Heisenberg chain \cite{shu16}. Then one can really identify out-of-phase dimerized
chain segments separated by localized spinons, and these spinons produce an RS state through interactions mediated by the gapped host segments. Similarly,
when starting from the VBS ordered 2D $J$-$Q$ model, localized spinons induced by randomness can be identified which appear pairwise and interact with each
other mainly through the domain walls \cite{luliu} and form a critical state. It is not clear why the spin correlations in this state decay as $r^{-2}$,
which is also the form in the 1D RS state, but a power-law decay in any case is a manifestation of physics similar to the 1D RS state.

In the system studied in the present paper, the dynamic exponent $z\approx 4$ and it was previously know that it can grow much larger---up to $z \approx 7$
was found at points deeper inside the RS phase in Ref.~\onlinecite{luliu} (and there is likely no upper bound in principle if other suitable interactions
are considered). Thus, the system can exhibit slow dynamics. With the
imposed exponent relation $\eta=2-z$, we found that the on-site (imaginary time) dynamic spin-spin correlation function takes the asymptotic form
$\langle {\bf S}_i(\tau){\bf S}_i(0)\rangle \propto \tau^{-2/z}$, representing a slow decay when $z$ becomes large. Here it should be noted that the slow
dynamics should correspond to weak effective couplings between spins in long valence bonds, in analogy with the 1D RS state where the logarithmic form of
the asymptotic decay of the correlations originates from rare very long bonds. Thus, while the finite-$z$ 2D state does not have the extreme slow dynamics
of the 1D RS system, it can still approach that behavior when $z$ becomes large, and for the same physical reasons.

Given all the above reasoning, the term 2D RS state for the critical phase in the random $J$-$Q$ model appears appropriate, despite the fact that it does not
correspond to an IRFP. It is not even clear whether an IRFP exists in SU($2$) spin models with realistic short-range interactions,
given the many negative results based on the SDRG method \cite{rnbhatt,melin00,damle00,motrunich00,lin03,lin06,laumann12,kimchinatcom}. These studies have also
found finite-disorder fixed points, though they were not characterized in the kind of detail achieved here, e.g., as regards the $r^{-2}$ form of the spin
correlations.

Finally, with regards to terminology we note that Griffiths phases \cite{griffiths69} are gapless phases of random systems appearing in the neighborhood
of transitions into ordered phases. However, physics involved in the formation of a quantum Griffths phase is quite different from the mechanisms underlying
the RS state. The gapless behaviors arise from rare, arbitrarily large ordered clusters within an otherwise gapped bulk. Such ordered clusters do not appear
in the RS phase, where the critical properties are intrinsic to the network of interacting spinons. The differences are well illustrated by dilute magnetic
semiconductors, where random antiferromagnetic interactions lead to RS physics \cite{rnbhatt}, and there is not necessarily any ordered phase close by which
spawns the RS state. In contrast, in the neighborhood of a ferromagnetic transition, a Griffiths phase with completely different properties arises
\cite{galitski04} due to the presence of large ferromagnetic clusters. The properties of a Griffiths phases in general depend sensitively on the nearby
phases which it is a mixture of. A quantum Griffiths phase has been discussed in the content of spin ice, where the classical ice state changes character
in the presence of strong quantum fluctuations and becomes a Griffiths phase (referred to as a disorder-induced spin liquid), located in parameter space
between a paramagnetic phase and a Coulomb spin liquid \cite{savary17}. Again, this Griffiths phase is very different from the RS phase. The RS state
with very specific properties discussed here should not be confused with a generic Griffiths state.

\subsection{Generality of the 2D RS state}
\label{sub:general}

While we have not proved that the RS state studied here is also generic beyond the ``designer'' interactions of the $J$-$Q$ model, there is mounting
evidence for this being the case. Recently, a DMRG study  \cite{huanda20} of the square-lattice $S=1/2$ Heisenberg model with random frustrated bonds
detected spin correlation functions apparently decaying with the same power-law $r^{-2}$ as in the random $J$-$Q$ model. The systems accessible with DMRG,
beyond the very small lattice accessible with the ED method \cite{oitmaa01,watanabe14,kawamura14,shimokawa15,uematsu17,uematsu18,kawamura19,uematsu19,hanqing19},
are still not very large and the associated finite-$T$ properties of the system have not been studied yet. It should be possible to reach larger
systems with the DMRG method in the future, and progress has also been made on $T>0$ properties \cite{li18}.

From the physical perspective, Kimchi et al.~\cite{kimchiprx} also proposed that the mechanism causing the RS state in frustrated quantum magnets
should be VBS domains with topological defects carrying spinons, exactly as found in the random $J$-$Q$ model \cite{luliu}. If indeed the physical ingredients
are the same, and the spinons in the $J$-$Q$ model do not form antiferromagnetic long-range order (which was proposed \cite{kimchiprx} but has not
been observed), then there is no strong reason why frustrated interactions (in the conventional sense of inability to satisfy antiferromagnetic spin
orientations along any loop on the lattice) should cause a low-energy state different that resulting from the different forms of competing interactions
of the $J$-$Q$ model. In a recent work, Dey and Vojta \cite{dey20} also made a similar remark in the context of a semi-classical analysis of spin glasses
in weakly frustrated 2D systems, suggesting that strong quantum effects should induce local VBS formation and transform the quasi-classical spin glass
into the proposed RS state \cite{kimchiprx,luliu}. This picture as well does not appear to depend on what causes the VBS domains.

Kimchi et al.~\cite{kimchiprx} also speculated that the RS state in the frustrated systems may only be an intermediate-energy state, which could be unstable
to formation of a spin glass at the lowest energies. As we discussed in Sec.~\ref{sub:2dstates}, there are no concrete reliable calculations showing that
a spin glass state is even possible in 2D $S=1/2$ systems with short-range interactions; instead, the RS state may be what replaces a classical or
semi-classical spin glass when the quantum fluctuations are strong---this is also the picture discussed by Dey and Vojta \cite{dey20}.

Numerically we can in principle not completely rule out that the RS state we have characterized is unstable to antiferromagnetic order at even lower
temperatures than we have reached. However, the ordering scenario appears implausible  based on the consistent power-law scaling observed for all the quantities
studied, down to what already is a very low temperature scale (of the order $10^{-2}$ of the microscopic coupling scale). In contrast,
Kimchi et al.~\cite{kimchiprx} did not even discuss
the possibility of an intermediate RS flow in systems without conventional frustration, implicitly assuming that the ordering tendency is strong unless
frustrated interactions are present. At the very least, it is now clear that the RS state we have characterized exists over a vast range of length and
energy scales, and it must correspond to an RG fixed point of random 2D quantum magnets.

To better connect the picture of the frustrated RS state, SDRG calculations, and the prototypical RS state of the $J$-$Q$ model, it would be useful
to study the dynamic exponent within the SDRG scheme. Normally, the results of SDRG calculations are expressed in terms of an exponent $\gamma$,
which controls the low-temperature specific heat $C/T$ and susceptibility $\chi_{\rm u}$; they both scale as $T^{-\gamma}$ with $\gamma < 1$
\cite{rnbhatt,kimchinatcom}. In our picture, where the scaling is conventional quantum-critical scaling with the constraint $\eta=2-z$ (which
is important when considering the local susceptibility $\chi_{\rm loc}$, which attains the same divergence as $\chi_{\rm u}$ and $C/T$), the exponent
is expressed as $\gamma=1-2/z$. It would be interesting to extract the dynamic exponent in the SDRG procedure and test this relationship (which we
believe must apply). It would of course also be interesting to check the decay form of the spin-spin correlations within the SDRG framework, but unfortunately
it is difficult to extract real-space information from the decimation procedure in more than one dimension. As far as we are aware, this has never been done.

\subsection{Experiments}
\label{sub:experiments}

Many of the quantities we have studied here can be tested experimentally. The perhaps most fundamental and definite aspect of the RS state characterized
here is the $r^{-2}$ decay of the spin correlations in the ground state.
In principle this form can be tested by inelastic neutron scattering through the associated logarithmic divergence of
the static structure factor $S({\bf q})$ in the limit ${\bf q} \to (\pi,\pi)$. The static structure factor is obtained by integrating the dynamic structure
factor $S({\bf q},\omega)$ over $\omega$, but clearly the logarithmic divergence will not be easy to detect, given typical experimental resolution, temperature
effects, and impurity effects not corresponding solely to random couplings. Nevertheless, it may be useful to look for a possible anomalous shape of the peak
of $S({\bf q})$ versus ${\bf q}$ in candidate RS materials. The predicted linear dependence on $T^{-1}$ of the staggered susceptibility, Eq.~(\ref{stagtdep}),
may be more practical for tests with inelastic neutron scattering than $S(\pi,\pi)$; it can be obtained from the $\omega^{-1}$-weighted frequency integral
$S({\bf q},\omega)$ at ${\bf q}=(\pi,\pi)$.

The specific heat and the uniform magnetic susceptibility are more easily accessible than dynamic quantities,
but both may also be sensitive to effects beyond the disorder giving rise
to the RS state. In particular, impurity moments in addition to random couplings should lead to other contributions to the susceptibility, though anomalous,
non-Curie behaviors may still apply \cite{riedl19}.

In our previous work \cite{luliu}, we pointed out that ${\rm Sr_2CuTe_{1-x}W_xO_6}$, which for $x=0$ is a good realization of the square-lattice
$S=1/2$ Heisenberg model \cite{koga16,babkevich16}, is a promising candidate for RS physics. For W fractions $0 < x < 1$, the system is randomly
frustrated, with first- and second-neighbor Heisenberg couplings mediated by plaquette-centered Te or W ions, respectively. The system has columnar
antiferromagnetic order for $x \agt 0.7$ \cite{vasala14,walker16} but no order for $0.1 \alt x \alt 0.7$ \cite{mustonen18,watanabe18}.
Though a Curie tail was reported in the susceptibility \cite{watanabe18}, the data at low temperatures can actually be better fitted to power
law $\chi_{\rm u} \propto T^{-\gamma}$ with $\gamma \approx 0.7$ (weakly dependent on $x$ in the range where there is no magnetic order) \cite{luliu}.
On the one hand this behavior may indicate an RS phase, but on the other hand the specific heat
does not show the same power law $C/T \propto T^{-\gamma}$ that we have demonstrated here (and which holds also within the SDRG method \cite{kimchinatcom});
instead a shoulder anomaly is present in $C/T$ near $1.2$ K, followed at lower temperatures by a drop more rapid than the expected power
law \cite{watanabe18}. It is not clear, however, what role subtraction of claimed impurity contributions play here, as those ``impurities'' may at
least partially be the spinons of the RS state.

Kimchi et el.~suggested that YbMgGaO$_4$ is a likely RS system \cite{kimchiprx}. This material had previously been regarded as quantum spin liquid \cite{li15},
and later as a spin glass \cite{ma18}. Indeed the material exhibits a power-law decay of the specific heat and a divergent susceptibility \cite{li15,ma18}.
While the former is natural under some scenarios in spin liquids, the divergent susceptibility is not. The divergent susceptibility was previously attributed
to magnetic impurities not part of the collective bulk state  \cite{jshelton}. It would be interesting to further analyze the exponents and test whether the
specific heat and the susceptibility are mutually consistent in the way tested here for the $J$-$Q$ model.

A strong case for RS physics in LiZn$_2$Mo$_3$O$_8$ and related materials was made by Kimchi et al.~\cite{kimchinatcom}, primarily based on the behavior
of the system in a magnetic field and scaling forms obtained from the SDRG scheme and related considerations. We have not yet studied the random $J$-$Q$ model
in a magnetic field, but such calculations are possible \cite{adam1,adam2} and would be interesting to carry out in the future. LiZn$_2$Mo$_3$O$_8$ had
previously been proposed to realize a VBS state with $1/3$ of the spins remaining paramagnetic and subject to weak antiferromagnetic couplings \cite{jpsheck}.
Kimchi et al.~\cite{kimchinatcom} also proposed that H$_3$LiIr$_2$O$_6$ may realize the RS phase. At zero field, both $C/T$ and $\chi_{\rm u}$ have an low-temperature
form $T^{-\gamma}$ with $\gamma \approx 0.5$ \cite{hliiro}, and this behavior was originally attributed to a low density of spin defects in a bulk quantum spin
liquid. A complication with LiZn$_2$Mo$_3$O$_8$ and H$_3$LiIr$_2$O$_6$ is strong Dzyaloshinskii-Moriya interactions, which was taken into account in the treatment
by Kimchi et al.~\cite{kimchinatcom}. It is not clear whether the RS state as realized in the $J$-$Q$ model can capture all aspects of random quantum magnets when
additional effects, e.g., Dzyaloshinskii-Moriya interactions, are important. However, if the system still remains critical with finite dynamic exponent the general
scaling laws tested here should still hold, though not necessarily with the $\eta=2-z$ constraint.

Finally, we mention $\alpha$-Ru$_{1-x}$Ir$_x$Cl$_3$, which recently has attracted attention as a possible randomness-induced spin liquid
\cite{kelley17,shdo18,shdo20} or RS state \cite{baek20} when $x$ slightly exceeds $0.2$. While various power-law behaviors have been observed,
the exponents governing $\chi_{\rm u}$ and $C/T$ are distinctly different, and therefore the state appears to be quite different from the
conventional RS state. One possible explanation for the violations is that the system has both bond disorder and spin vacancies. In the
$J$-$Q$ model, vacancies induce long-range antiferromagnetic order because the spinon pairing imposed by the topological defects
can no longer be strictly maintained when the moments associated with vacancies are completely randomly distributed
\cite{luliu}. It is possible that some power-law scaling can still be observed in some range of temperatures, and in principle this can be tested
within the $J$-$Q$ model, for which the previous work only established that some long-range order is present. This long-range order can in principle
be destroyed in the presence of frustrated interactions, and then an RS state may still form. The effects of vacancies in a frustrated triangular-lattice
quantum magnet were theoretically studied by Riedl et al. \cite{riedl19}, but no concrete predictions for thermodynamics were presented. It would
be interesting to further study the interplay of vacancies and coupling disorder.

\begin{acknowledgments}
W.G. is supported by the NSFC under Grants No.~11734002 and No.~11775021. A.W.S. is supported by the NSF under Grant No.~DMR-1710170 and by a Simons
Investigator Award. L.L. Would like to thank Boston University's Condensed Matter Visitors Program for support. We also acknowledge the Super Computing
Center of Beijing Normal University and Boston University's Research Computing Services for their support.
\end{acknowledgments}

\end{document}